\documentclass[1p]{elsarticle}

\usepackage{lineno,hyperref}
\usepackage{float}
\usepackage{slashed}
\usepackage{mathtools}
\usepackage{longtable}
\DeclarePairedDelimiter\bra{\langle}{\rvert}
\DeclarePairedDelimiter\ket{\lvert}{\rangle}
\DeclarePairedDelimiterX\braket[2]{\langle}{\rangle}{#1 \delimsize\vert #2}
\DeclarePairedDelimiterX\inner[2]{\langle}{\rangle}{#1,#2}

\begin{document}

\begin{frontmatter}

\title{Properties of bottomonium in a relativistic quark model}


\author[mymainaddress]{Manjunath Bhat}

\author[mymainaddress]{Antony Prakash Monteiro\corref{mycorrespondingauthor}}
\cortext[mycorrespondingauthor]{Corresponding author}
\ead{aprakashmonteiro@gmail.com}

\author[mysecondaryaddress]{K. B. Vijaya Kumar}
\address[mymainaddress]{P. G. Department of Physics, St Philomena college
              Darbe, Puttur  574 202, India}
\address[mysecondaryaddress]{Department of Physics, Mangalore University,
Mangalagangothri P.O., Mangalore - 574199, INDIA}

\begin{abstract}
The mass spectrum of $b\bar{b}$ states has been obtained using the phenomenological relativistic quark model
(RQM). The Hamiltonian used in the investigation has  confinement potential and confined one gluon exchange potential (COGEP). In the frame work of RQM a study of M1 and E1 radiative decays of $b\bar{b}$ states have been made. The weak decay widths in the spectator quark approximation have been estimated. An overall agreement is obtained with the experimental masses and decay widths. 
\end{abstract}


\end{frontmatter}

\section{Introducing Bottomonium states}

\label{sec:intro}
  The bound state of a bottom quark b and its anti quark $\bar{b}$ known as bottomonium, play an important role in the study of the strong interactions. The spectrum of bottomonium is known very well from the experiment and there are many different approaches to calculate it theoretically. Phenomenological potential models provide an accurate way of doing this. Bottomonium was first observed in the spectrum of $\mu\mu^-$ pairs produced in 400 GeV proton$-$nucleus collisions at Fermilab. The $b\bar{b}$ states were dicovered as spin triplet states $\Upsilon(1S)$, $\Upsilon(2S)$ and $\Upsilon(3S)$ by E288 collaboration at Fermilab \cite{HS77,WR77}.  \\
  
  The spin-singlet state $\eta_b(1S)$ was discovered by the Babar Collaboration and they measured the mass of $\eta_b(1S)$ to be $M = 9388.9^{+3.1}_{-2.3}(stat)\pm  2.7(syst)$ MeV\cite{AB08}. In an another measurement the BaBar found the mass of $\eta_b(1S)$ to be $M=9394^{+4.8}_{-2.3}\pm 2.7$ MeV. The CLEO collaboration  measured the mass of $\eta_b(1S)$ to be $9391\pm 6.6$ MeV\cite{BG10}. The more precise measurement of mass of $\eta_b(1S)$ state is done by the Belle  collaboration  which obtained a value of $M=9402.4\pm 1.5\pm1.8$ MeV\cite{MR12}. The $\eta_b(2S)$ was successfully observed by the CLEO  collaboration  in $\Upsilon(2S)\to\eta_b(2S)\gamma$ decays at a mass of $9974.6\pm 2.3\pm 2.1$ MeV\cite{SD12}. The Belle collaboration  has reported a signal for $\eta_b(2S)$ using the $h_b(2P)\to\eta_b(2S)\gamma$ at a mass of $9999.0\pm 3.5^{+2.8}_{-1.9}$ MeV\cite{MR12}.\\
  
   The BaBar collaboration first reported the evidence for the spin singlet P wave state $h_b(1P)$ in the transition $\Upsilon(3S)\to \pi^0h_b(1P)\to\pi^0\gamma\eta_b(1S)$ \cite{LJ12}. Later the Belle collaboration found $h_b(1P)$ through $\Upsilon(5S)\to h_b(1P)\pi^+\pi^-$ transition \cite{AI12}. The measured masses of $h_b(1P)$ and $h_b(2P)$ were $9898.25\pm 1.06^{+1.03}_{-1.07}$ MeV and $10259.76\pm 0.64^{+1.43}_{-1.03}$ MeV respectively. The two triplet P-wave states $\chi_{bJ}(2P)$ and $\chi_{bj}(1P)$ with J=0,1,2 were discovered in radiative decays of the $\Upsilon(3S)$ and $\Upsilon(2S)$ respectively \cite{KH83,EG82,CK83, FH83}. The states  $\chi_{bJ}(nP)$ were produced in the proton - proton collisions at the LHC at $\sqrt{s}=7$ TeV and recorded by the ATLAS detector\cite{GA12}. In addition to this, a new state $\chi_{bJ}(3P)$ centred at a mass of $10530\pm 5\pm 9$ MeV has been observed in both the $\Upsilon(1S)\gamma$ and $\Upsilon(2S)\gamma$ decay modes. This state was confirmed by the D0 collaboration which observed the $\chi_{bJ}(3P)$ state in the $\Upsilon(1S)\gamma$ final state with mass of $10551\pm 14\pm 17$ MeV\cite{VM12}. The LHCb collaboration has measured the mass of the $\chi_{b1}(3P)$ to be $10515^{+2.2}_{-3.9}(\rm{stat})^{+1.5}_{-2.1}(\rm{syst})$ MeV\cite{AR14}. \\
   
   The spin triplet D-wave state $\Upsilon(1^3D_2)$ was observed through the decay chain $\Upsilon(3S)\to\gamma\gamma\Upsilon(1^3D_2)\to\gamma\gamma\pi^+\pi^-\Upsilon(1S)$\cite{GB04,PD10}.  The $\Upsilon(^3D_J)$ was discovered in the $\pi^+\pi^-\Upsilon(1S)$ final state with mass $M = 10164.5\pm 0.8(stat)\pm 0.5(syst)$ MeV\cite{PD10}. The radial excitations of vector bottomonium family $\Upsilon(4S)$, $\Upsilon(10860)$ and $\Upsilon(11020)$ were also observed\cite{BD85,LD85}. The two states $X(10610)^{\pm}$ and $X(10650)^\pm$ were observed by the Belle collaboration in the mass spectra of $\pi^\pm\Upsilon(nS)$ (n=1,2,3) and $\pi^\pm h_b(nP)$ (n=1,2) pairs\cite{AB12}. These states do not fall into the usual scheme of mesons since they posses electrical charge. These states can be identified as molecular states, since these are very near to the $BB^*$ and $B^*B^*$ thresholds.\\
 
 QCD motivated potential models have played an important role in the understanding of quarkonium spectroscopy. These models have successfully  predicted the bottomonium spectrum but over estimated the decay rates.  In the present work we study bottomonium spectrum in detail using relativistic quark model. \\

The paper is organized in 4 sections. In sec.~\ref{sec:TB} we briefly review the theoretical background for relativistic model and the relativistic description of radiative decay widths. In sec.~\ref{sec:RD} we discuss the results and the conclusions are drawn in sec.~\ref{sec:C} with a comparison to other models. \\ 

\section{Phenomenological Understanding}

\label{sec:TB}
\subsection{The Relativistic Harmonic Model}

Describing a bottomonium bound state problem, incorporating the elements of non perturbative low energy dynamics is a challenging problem. The two main approaches in this line have been effective field theories\cite{HG90,M97,NB05} and lattice gauge theories. Though the calculations away from threshold have been proven successful, the threshold region remain still tricky. And hence a relativistic potential model would be a befitting one for our study of bottomonium states. Essentially in all phenomenological relativistic QCD base quark models, the Hamiltonian for the quark system consists of a two body confinement potential and Coulomb like potential which takes care of relativistic effects. The Hamiltonian in our model  has the confinement potential and a two body confined one gluon exchange potential(COGEP) \cite{PC92, SB91,KB93, KB97}. \\
The RHM equation  is 
 \cite{SB83,VH04,VB09}:

\begin{equation}
 (\alpha \cdot p + \gamma_{0} M + \frac{1}{2} (1+ \gamma_{0})A^{2} r^{2})\psi= E \psi  
\end{equation}  

The Dirac wave function $\psi$ is written as $   \mid \psi  \rangle = N  \left( \begin{array}{cc}
\phi \\ 
\chi \end{array} \right) $,where $ \phi$ is the large componet and $\chi$ is the small component for positive energy state solution in the non-relativistic limit. Using standard matrix form of $\alpha$ and $\beta$,the Dirac equation is expressed in terms of the large and small components.

\begin{equation}
 (A^{2} r^{2} +M) \phi +( \sigma \cdot p) \chi = E \phi 
\end{equation}
\begin{equation} (\sigma \cdot p) \phi -(M+E) \chi =0 
\end{equation} 

 Hence, solution of the full Dirac equation   $\psi$ is written as 
\begin{equation}
   \mid \psi  \rangle = N  \left( \begin{array}{cc}
\phi \\
\frac {\sigma \cdot p}{E+M} \phi
 \end{array} \right) 
\end{equation}

where $\gamma_0$ is the Dirac matrix, M is a constant mass and $A^2$ is the confinement strength. \\

with normalization constant 
\begin{equation}
N=\sqrt{\frac{2(E+M)}{3E+M}}\label{n1}
\end{equation}
E is the eigenvalue of the single particle Dirac equation with interaction potential given by equation~(\ref{eq:A}). We perform a unitary transformation to eliminate the lower component of $\psi$. With this transformation, $\phi$ satisfies the 'harmonic oscillator' wave equation
\begin{equation}
\left(\frac{p^2}{E+M}+A^2r^2\right)\phi=(E-M)\phi \label{ha}
\end{equation}

the eigenvalue of which is given by
\begin{equation}
E^2_N=M^2+(2N_{OSC}+1)\Omega_N
\end{equation}
where $N_{OSC}=2n+l$ is the oscillator quantum number and $\Omega_N$ is the energy dependent oscillator size parameter given by
\begin{equation}
\Omega_N=A(E_N+M)^{1/2}
\end{equation}
The total energy or the mass of the meson is obtained by adding the individual contributions of the quarks. The spurious centre of mass (CM) is corrected \cite{GG57} by using intrinsic operators for the $\sum_ir^2_i$ and $\sum_i\nabla^2_i$ terms appearing in the Hamiltonian. This amounts to just subtracting the CM motion zero contribution from the $E^2$ expression. 
\subsection{Confined One Gluon Exchange Potential}
 NRQM employing OGEP, which takes into consideration the confining effect of quarks on mesonic states, does not shed light on confinement of gluons. Hence one needs to incorporate confining effects of gluons on mesonic states since confined dynamics of gluons plays a decisive role in determining the hadron spectrum and in hadron-hadron interaction. There are various confinement models for the gluons. The effect of the confined gluons on the masses of mesons has been studied using the successful current confinement model (CCM). The confined gluon propagators (CGP) are derived in CCM to obtain the COGEP.\\

 The central part of the COGEP is\cite{PC92, SB91,KB93, KB97}
\begin{equation}
V^{cent}_{COGEP}(\vec{r})=\frac{\alpha_sN^4}{4}\lambda_i\cdot\lambda_j\left[D_0(\vec{r})+\frac{1}{(E+M)^2}\left[4\pi\delta^3(\vec{r})-c^4r^2D_1(\vec{r})\right]\left[1-\frac{2}{3}\vec{\sigma}_i\cdot\vec{\sigma}_j\right]\right]
\end{equation}
where $\lambda_i$, $\lambda_j$ are the color matrices, $D_0(\vec{r})$ and $D_1(\vec{r})$ are the propagators given by
\begin{eqnarray}
D_0(\vec{r})=\frac{\Gamma_{1/2}}{4\pi^{3/2}}c(cr)^{-3/2}W_{1/2;-1/4}(c^2r^2)\\
D_1(\vec{r})=\frac{\Gamma_{1/2}}{4\pi^{3/2}}c(cr)^{-3/2}W_{0;-1/4}(c^2r^2)
\end{eqnarray}
Here $\Gamma_{1/2}=\sqrt{\pi}$, W's are Whittaker functions and $c$(fm$^{-1}$) is a constant parameter which gives the range of propagation of gluons and is fitted in the CCM to obtain the glue-ball spectra and r is the distance from the confinement center.\\ 

The spin orbit part of COGEP is
\begin{equation}
\begin{split}
V^{LS}_{12}(\vec{r})=\frac{\alpha_s}{4}\frac{N^4}{(E+M)^2}\frac{\lambda_1\cdot\lambda_2}{2r}
\times\left[[\vec{r}\times(\hat{P}_1-\hat{P}_2)\cdot(\sigma_1+\sigma_2)](D'_0(\vec{r})+2D'_1(\vec{r}))+\right.\\
\left.[\vec{r}\times(\hat{P}_1+\hat{P}_2)\cdot(\sigma_1-\sigma_2)](D'_0(\vec{r})-D'_1(\vec{r}))\right]
\end{split}
\end{equation}
where $\hat{P}_1$, $\hat{P}_2$ are the momentum of quarks.
The spin orbit term has been split into the symmetric $(\sigma_1+\sigma_2)$ and anti symmetric $(\sigma_1-\sigma_2)$ spin orbit terms.\\

The tensor part of the COGEP is,
\begin{equation}
V^{TEN}_{12}(\vec{r})=-\frac{\alpha_s}{4}\frac{N^4}{(E+M)^2}\lambda_1\cdot\lambda_2\left[\frac{D''_1(\vec{r})}{3}-\frac{D'_1(\vec{r})}{3r}\right]S_{12}
\end{equation}
where
 \begin{equation}
 S_{12}=[3(\sigma_1\cdot\hat{r})(\sigma_2\cdot\hat{r})-\sigma_1\cdot\sigma_2]
 \end{equation}
The total potential is then 
\begin{equation}
\label{pot}
V(\vec{r})=V^{cent}_{COGEP}(\vec{r})+V^{LS}_{12}(\vec{r})+V^{TEN}_{12}(\vec{r})
\end{equation} 
We then obtain Hamiltonian by adding interaction potential given by equation \ref{pot} to the the 'harmonic oscillator' wave equation  \ref{ha},
\begin{equation}
H=\sum_{i=1}^2\frac{P^2_i}{E+M}-K_{CM}-\sum_{i<j}A^2r^2_{ij}\lambda_i\cdot\lambda_j+V(\vec{r})
\end{equation}
\subsection{Radiative Decays}
We consider two types of radiative transitions of the $b\bar{b}$ meson:
\subsubsection{Electric Dipole (E1) Transitions} 
The partial widths for electric dipole (E1) transitions between states $^3S_1$ and $^3P_J$ are given by
\begin{equation}
\Gamma_{(i\to f+\gamma)}=(2J'+1)\frac{4}{3}Q^2_b\alpha k^3_0S_{if}^E\left|{\cal E}_{if}\right|^2 
\end{equation}

where $k_0$ is the energy of the emitted photon,\\

~~~~~~~~~~~~~~$k_0=\frac{m^2_i-m^2_f}{2m_i}$ in relativistic model.\\

$\alpha$ is the fine structure constant. $Q_b=1/3$ is the charge of the b quark in units of $|e|$. $m_i$ and $m_f$ are the masses of initial and final mesons. The statistical factor $S_{if}^E=\rm{max}(l,l')\left \{\begin{array}{ccc}
J & 1 & J' \\
l' & s  & l 
 \end{array} \right\}^2$, 
$J,~J'$ are the total angular momentum of initial and final mesons, $l,~l'$ are the orbital angular momentum of initial and final mesons and $s$ is the spin of initial meson.
 \begin{equation}
 {\cal E}_{if}=\frac{3}{k_0}\int^\infty_0 r^3R_{nl}(r)R'_{nl}(r)dr\left[\frac{k_0r}{2}j_0\left(\frac{k_0r}{2}\right)-j_1\left(\frac{k_0r}{2}\right)\right]
 \end{equation}
 is the radial overlap integral which has the dimension of length, with $R_{nl}(r)$ being the normalized radial wave functions for the corresponding states.\\

\subsubsection{Magnetic Dipole (M1) Transitions}
The M1 partial decay width between S wave states is \cite{WJ88,NL78,CE05,WE86,BB95,NB99,BA96}
\begin{equation}
\label{decay}
\begin{split}
\Gamma_{i\to f\gamma}=\delta_{L L'}4\alpha k^3_0\frac{E_b(k_0)}{m_i}\left(\frac{Q_b}{m_b}+(-1)^{S +S'}\frac{Q_{\bar{b}}}{m_{\bar{b}}}\right)^2(2S+1)
\times(2S'+1)(2J'+1)\\
\left \{\begin{array}{ccc}
S & L & J \\
J' & 1  & S' 
 \end{array} \right\}^2\left \{\begin{array}{ccc}
1 & \frac{1}{2} & \frac{1}{2} \\
\frac{1}{2} & S  & S' 
 \end{array} \right\}^2
\times\left[\int^\infty_0  R_{n'L'}(r)r^2j_0(kr/2)R_{nL}(r) dr\right]^2
\end{split}
\end{equation} 
where $\int^\infty_0 dr  R_{n'L'}(r)r^2j_0(kr/2)R_{nL}(r)$ is the overlap integral for unit operator between the coordinate wave functions of the initial and the final meson states, $j_0(kr/2)$ is the spherical Bessel function, $m_b$ is the mass of bottom quark. $J$ and $J'$ are the total angular momentum of initial and final meson states respectively. $L$, $L'$, $S$ and $S'$ are the orbital angular momentum ans spin angular momentum of initial and final meson states respectively \\

\section{Discussion on Results}
\label{sec:RD}
\subsection{Mass Spectra}

The quark-antiquark wave functions in terms of oscillator wave functions corresponding to the relative and center of mass coordinates have been expressed here, which are of the form, 
\begin{equation}
\Psi_{nlm}(r,\theta, \phi) = N' (\frac{r}{b})^{l}~L_{n}^{l+\frac{1}{2}}(\frac {r^2}{b^2})\exp(-\frac{r^2}{2b^2})Y_{lm}(\theta, \phi)
\end{equation}
where $N'$ is the normalising constant given by 
\begin{eqnarray}
{\lvert N' \rvert} ^2={\frac{2n!}{b^3 \pi^{1/2}}} \frac{2^{[2(n+l)+1]}}{(2n+2l+1)!}(n+l)!
\end{eqnarray} 
$L_{n}^{l+\frac{1}{2}}$ are the associated Laguerre polynomials. \\

The harmonic oscillator wave function  allows the separation of the motion of the center of mass and has been widely used to classify the spectra of baryons and mesons \cite{FD69,RP71} and  extending to nucleon-nucleon interaction is straight forward \cite{KS89,FA83,KB93}.  If the basic states are the harmonic oscillator wave functions, then it is straightforward to evaluate the matrix elements of few body systems such as mesons or baryons. Since the basic states are the products of the harmonic oscillator wave functions they can be chosen in a manner that allows the product wave functions to be expanded as a finite sum of the corresponding products for any other set of Jacobi coordinates.  
It is advantageous to use the Gaussian form since in the annihilation of quark-anti quark into lepton pairs, the amplitude of the emission or absorption processes depends essentially on the overlap of initial and final hadrons and hence the overlap depends only on the intermediate  distance region of the spatial wave functions  which can extend up to 0.5 fm. This intermediate region can be described by potentials that are similar in this region and hence harmonic oscillator wave functions are expected to reproduce emission and absorption processes quite well.\\

The five parameters used in our model are  the mass of beauty quark $m_b$, the harmonic oscillator size parameter $b$, the confinement strength $A^2$, the CCM parameter $c$ and the quark-gluon coupling constant $\alpha_{s}$. There are several papers in literature where the size parameter $b$ is defined \cite{SN86,IM92}. The value of $b$ is fixed by minimizing the expectation value of the Hamiltonian for the vector meson. To start with, we construct the $5\times 5$ Hamiltonian matrix for $b\bar{b}$ states in the harmonic oscillator basis. The confinement strength $A^2$ is fixed by the stability condition for variation of mass of the meson against the size parameter $b$. 
$$\frac{\partial}{\partial b}\bra{\psi}H\ket{\psi}=0$$

The parameter $c$ in CCM \cite{PC92,SB85,SB87} was obtained by fitting the iota (1440 MeV)$0^{-+}$ as a digluon glue ball.To fit $\alpha_s$ , $m_b$  we start with a set of reasonable values and diagonalize the matrix for $b\bar{b}$ meson. Then we tune these parameters to obtain an agreement with the experimental value for the mass of $b\bar{b}$ meson. In literature we find different sets of values for $m_b$, which are listed in Table \ref{mass1}.\\
\begin{table}[!h]
\centering
\caption{\label{mass1}\bf m$_b$ for various theoretical models (in MeV).}
\setlength{\tabcolsep}{2pt}
\begin{tabular}{ccccccc}
\hline
Parameter&Ref.\cite{AM80}&Ref. \cite{EJ78}&Ref. \cite{WS81}& Ref. \cite{CJ77}& Ref.\cite{DR03}\\
\hline
$m_b$&5174 &5180 &4880 &4880 &4880 \\
\hline
\end{tabular}
\end{table}

The values of strong coupling constant $\alpha_s$ in literature are listed in Table \ref{alpha}. The value of strong coupling constant ($\alpha_{s}$) used is compatible with the perturbative treatment.\\
\begin{table}[!h]
\centering
\caption{\bf $\alpha_s$ for various theoretical models.}
\label{alpha}
\setlength{\tabcolsep}{2pt}
\begin{tabular}{ccccccc}
\hline
Parameter&Ref. \cite{SN85}&Ref. \cite{DR03}&Ref.\cite{AA05}& Ref. \cite{EC94}& Ref. \cite{SA95}\\
\hline
$\alpha_s$& 0.21&0.265&0.357&0.361&0.391\\
\hline
\end{tabular}
\end{table}
 The mass spectrum has been obtained by diagonalizing the Hamiltonian in a large basis of $5\times 5$ matrix. The calculation clearly indicates that masses for both pseudo scalar and vector mesons converge to the experimental values when the diagonalization is carried out in a larger basis. In our earlier work also, we had come to the similar conclusion while investigating light meson spectrum\cite{VH04,BK05}. The diagonalization of the Hamiltonian matrix in a larger basis leads to the lowering of the masses and justifies the perturbative technique to calculate the mass spectrum. The calculation clearly indicates that when diagonalization is carried out in a larger basis convergence is achieved both for pseudo-scalar mesons and vector mesons to the respective experimental values. \\
We use the following set of parameter values.
\begin{equation}
\begin{split}
m_b=4770.0~{\rm MeV}~~;b
= 0.35~{\rm fm};~~~ \alpha_s = 0.2~{\rm{to}}~0.3;~~A^2=220~{\rm MeV~fm^{-2}};~~c=1.72~{\rm fm^{-1}}
\end{split}
\end{equation}

The calculated masses of the $b\bar{b}$ states after diagonalization are listed in Table ~\ref{spectrum}. Our calculated mass value for $\eta_b(1S)$ is 9399 MeV which agrees with the experimental value 9398.0$\pm 3.2$ MeV\cite{PDG} and for $\Upsilon$(1S) is 9460.30 $\pm 0.26$. $J/\psi$(1S) is heavier than $\eta_b(1S)$ by 61 MeV. This difference is justified by calculating the $^3S_1-{}^1S_0$ splitting of the ground state which is given by 
 \begin{equation}
 M({}^3S_1)-M({}^1S_0)=\frac{32\pi\alpha_s|\psi(0)|^2}{9m^2_b}
 \end{equation}
 The experimental value of hyperfine mass splitting for the ground state is given by \cite{PDG}
 \begin{equation}
 M({}^3S_1)-M({}^1S_0)=62.3\pm 3.2~\rm{MeV}
 \end{equation}
The hyperfine mass splitting calculated in our model is in good agreement with both experimental data\cite{PDG} and lattice QCD result ($60.3\pm7.7)$ MeV\cite{SM10}. The  EFTs have predicted a hyperfine mass splitting of $41\pm 11^{+9}_{-8}$ MeV\cite{BA04} which is lower than the experimental value.\\

We predict a mass of 10001.12 MeV for the $\eta_b(2S)$ state which is in good agreement with the experemental value $9999.0\pm 3.5$MeV \cite{PDG}. The calculated mass of spin triplet state($\Upsilon(2S)$) of $\eta_b(2S)$ is 10024.50 MeV which is found to be in good agreement with the experemntal value $10023.26\pm 0.31$ MeV. The hyperfine mass splitting for the 2S state $\Delta_{hf}(\eta_b(2S))=23.38$ MeV is in good agreement with both the experemental one $\Delta_{hf}(\eta_b(2S))=24.3^{+4.0}_{-4.5}$ MeV\cite{PDG} and with the lattice QCD results $\Delta_{hf}(\eta_b(2S))=(23.5-28.0)$ MeV\cite{SM10}. The predicted mass values for $\eta_b(3S)$ and $\Upsilon(3S)$ states in our model are 10375.96 MeV and 10392.00 MeV respectively. The corresponding hyperfine mass splitting is $\Delta_{hf}(\eta_b(3S))=16.04$ MeV. We have also claculated masses of higher radially excited states of $\eta_b$ and its spin triplet state $\Upsilon$.  
  The mass of spin singlet P wave state $h_b(1P)$ calculated in our model is compatible with the experimental one. The masses of spin triplet states $\chi_{bJ}$ are in good agreement with experimental values of masses of these states. Our prediction for masses of higher excited P wave states are 10-20 MeV higher than the experimental values and other model calculations. Some of the higher excited states are 50-100 MeV heavier in our model.
\begin{table*}[h!]
\centering
\caption{\label{spectrum}\bf Bottomonium  mass spectrum (in MeV).}
\begin{tabular*}{\textwidth}{@{\extracolsep{\fill}}lrrrrrrrrrrrrrrrl@{}}
\hline
State &&\\
$n~^{2S+1}L_J$&\multicolumn{1}{c}{This work}&\multicolumn{1}{c}{Experiment\cite{PDG}}&\multicolumn{1}{c}{Ref.\cite{JS15}}&\multicolumn{1}{c}{Ref. \cite{SG16}}&\multicolumn{1}{c}{Ref. \cite{JF13}}&\multicolumn{1}{c}{Ref. \cite{Li12}}&\multicolumn{1}{c}{Ref.\cite{YL16}}\\
\hline
$1~^1S_0$&9399.74&9398.0$\pm$3.2&9455&9402&9391&9391.8&9394.5\\
$1~^3S_1$&9460.77&9460.30$\pm$0.26&9502&9465&9489&9460.3&9459.2\\
$1^3P_0$&9856.75&9859.44$\pm$0.42&9855&9847&9849&9875.3&9850.0\\
$1^1P_1$&9898.95&9899.3$\pm$1.0&9879&9882&9885&9915.5&9884.4\\
$1^3P_1$&9903.73&9892.78$\pm$0.26&9874&9876&9879&9906.8&9878.4\\
$1^3P_2$&9919.21&9912.21$\pm$0.26&9886&9897&9900&9929.6&9897.1\\
$2~^1S_0$&10001.12&9999.0$\pm$3.5&9990&9976&9980&10004.9&9982.6\\
$2~^3S_1$&10024.50&10023.26$\pm$0.31&10015&10003&10022&10026.2&10012.1\\
$1^3D_1$&10151.14&-&10117&10138&10112&10138.1&10132.8\\
$1^1D_2$&10160.69&-&10123&10148&10122&10145.5&-\\
$1^3D_2$&10163.48&-&10122&10147&10121&10144.6&10138.3\\
$1^3D_3$&10170.97&-&10127&10155&10127&10149.3&-\\
$2^3P_0$&10244.15&10232.5$\pm$0.4&10221&10226&10226&10227.9&10233.6\\
$2^1P_1$&10269.15&10259.8$\pm$0.5&10240&10250&10247&10259.1&10262.7\\
$2^3P_1$&10275.77&10255.46$\pm$0.22&10236&10246&10244&10252.4&10257.7\\
$2^3P_2$&10287.14&10268.65$\pm$0.22&10246&10261&10257&10270.1&10273.9\\
$3~^1S_0$&10375.96&-&10330&10336&10338&10337.9&10358.6\\
$3~^3S_1$&10392.00&10355.2$\pm$0.5&10349&10354&10358&10351.9&10379.9\\
$2^3D_1$&10468.44&-&10414&10441&-&10420.4&10453.9\\
$2^1D_2$&10478.94&-&10419&10450&-&10427.98&-\\
$2^3D_2$&10479.40&-&10418&10449&-&-&-\\
$2^3D_3$&10486.20&-&10422&10455&-&-&-\\
$3^1P_1$&10560.66&-&10516&10541&10591&-&10568.5\\
$3^3P_0$&10566.17&-&10500&10522&10495&-&10540.2\\
$3^3P_1$&10589.96&10515.7$\pm$2.2&10513&10538&10580&-&10563.9\\
$3^3P_2$&10599.93&-&10521&10550&10578&-&10576.2\\
$4~^1S_0$&10692.09&-&-&10623&-&-&-\\
$4~^3S_1$&10704.51&10579.4$\pm$1.2&10607&10635&-&-&10718.0\\
$3^3D_1$&10757.06&-&10653&10698&-&-&10752.5\\
$3^1D_2$&10766.28&-&10658&10706&-&-&-\\
$3^3D_2$&10766.51&-&10657&10705&-&-&-\\
$3^3D_3$&10772.55&-&10660&10711&-&-&-\\
$4^1P_1$&10751.68&-&-&10790&-&-&-\\
$4^3P_0$&10857.83&-&-&10775&-&-&-\\
$4^3P_1$&10876.06&-&-&10788&&-&-&-\\
$4^3P_2$&10882.90&-&10744&10798&-&-&-\\
$5~^1S_0$&10980.94&-&-&10869&-&-&-\\   
$5~^3S_1$&10990.94&10876$\pm$11&10818&10878&-&-&10997.1\\
$4^3D_1$&11028.52&-&-&10928&-&-&11031.9\\
$4^1D_2$&11036.36&-&-&10935&-&-&-\\
$4^3D_2$&11036.38&-&-&10934&-&-&-\\
$4^3D_3$&11041.45&-&-&10939&-&-&-\\
$5^3P_0$&11133.06&-&-&10798&-&-&-\\
$5^3P_1$&11146.87&-&-&11004&-&-&-\\
$5^3P_2$&11153.78&-&-&11022&-&-&-\\
$5^1P_1$&10943.12&-&-&11016&-&-&-\\
$5^3D_1$&11287.23&-&-&-&-&-&11268.9\\
$5^1D_2$&11293.53&-&-&-&-&-&-\\
$5^3D_2$&11293.48&-&-&-&-&-&-\\
$5^3D_3$&11297.86&-&-&-&-&-&-\\
$6~^1S_0$&11255.33&-&-&11097&-&-&-\\   
$6~^3S_1$&11308.41&11019$\pm$8&10995&11102&-&-&11232.3\\
\hline
\end{tabular*}
\end{table*}
\subsection{Radiative Decays}
Radiative decays of excited bottomonium states are the powerful tools which can be used to study the internal structure of $b\bar{b}$ states and they provide a good test for the predictions from the various models. Radiative transitions between different bottomonium states can be used to discover new bottomonium states experimentally.\\

The possible $E1$ decay modes have been listed in Table \ref{E1} and the predictions for E1 decay widths are given. Also our predictions have been compared  with other theoretical models. Most of the predictions for $E1$ transitions are in qualitative agreement. However, there are some differences in the predictions  due to differences in phase space arising from different mass predictions and also from the wave function effects particularly from relativization procedure.  We find our results are compatible with other theoretical model values for most of the channels. Relativistic
corrections to the wave functions tend to reduce the E1 transition widths for most of the channels. We find that the E1 transitions $3S\to 1P$ are suppressed compared to other E1 transitions and it is interesting to see that the E1 transition rates for $3S\to 1P$ are zero in the nonrelativistic case\cite{EG08,GR92}. We find that these E1 transisitions are very sensitive to relativistic effects.\\

\begin{table*}[h!]
\centering
\caption{\label{E1}\bf E1 transition rates of bottomonium.}
\begin{tabular*}{\textwidth}{@{\extracolsep{\fill}}lrrrrrrrrl@{}}
\hline
Transition&\multicolumn{1}{c}{k$_0$}&\multicolumn{1}{c}{This work}&\multicolumn{1}{c}{Experiment\cite{PDG}}&\multicolumn{1}{c}{Ref. \cite{CE05}}&\multicolumn{1}{c}{Ref. \cite{SR07}}&\multicolumn{1}{c}{Ref. \cite{SG16}}&\multicolumn{1}{c}{Ref.\cite{JS15}}\\
&MeV&keV&keV&keV&keV&keV\\
\hline
$1^3P_0\rightarrow 1^3S_1\gamma$&388.02&22.046&-&22.1&22.1&23.8&28.07\\
$1^3P_1\rightarrow 1^3S_1\gamma$&433.05&29.754&-&27.8&27.3&29.5&35.66\\
$1^3P_2\rightarrow 1^3S_1\gamma$&447.846&32.567&-&31.6&31.2&32.8&39.15\\
$1^1P_1\rightarrow 1^1S_0\gamma$&428.481&28.913&-&41.8&37.9&35.7&43.66\\
$2^3S_1\rightarrow 1^3P_0\gamma$&166.346&1.248&1.21$\pm 0.16$&1.29&1.15&0.91&1.09\\
$2^3S_1\rightarrow 1^3P_1\gamma$&120.042&0.479&2.20$\pm 0.22$&2.0&1.87&1.63&1.84\\
$2^3S_1\rightarrow 1^3P_2\gamma$&104.73&0.320&2.28$\pm 0.22$&2.04&1.88&1.88&2.08\\
$2^1S_0\rightarrow 1^1P_1\gamma$&101.647&0.292&-&1.99&4.17&2.48&2.85\\
$2^3P_0\rightarrow 1^3S_1\gamma$&753.427&2.584&-&10.9&6.69&2.5&5.44\\
$2^3P_1\rightarrow 1^3S_1\gamma$&782.68&3.284&-&12.0&7.31&5.5&9.13\\
$2^3P_2\rightarrow 1^3S_1\gamma$&793.178&3.569&-&12.7&7.74&8.4&11.38\\
$2^1P_1\rightarrow 1^1S_0\gamma$&832.606&4.822&-&-&-&13.0&14.90\\
$2^3P_0\rightarrow 2^3S_1\gamma$&217.287&6.695&-&9.17&9.90&10.9&12.80\\
$2^3P_1\rightarrow 2^3S_1\gamma$&248.197&9.729&-&12.4&13.7&13.3&15.89\\
$2^3P_2\rightarrow 2^3S_1\gamma$&259.287&14.643&-&14.5&16.8&14.3&17.5\\
$2^1P_1\rightarrow 2^1S_0\gamma$&264.532&15.476&-&19.0&37.9&14.1&17.60\\
$3^3S_1\rightarrow 2^3P_0\gamma$&146.798&6.792&1.19$\pm  0.16$&1.35&1.67&1.03&1.21\\
$3^3S_1\rightarrow 2^3P_1\gamma$&115.580&3.385&2.56$\pm 0.34$&2.20&2.74&1.91&2.13\\
$3^3S_1\rightarrow 2^3P_2\gamma$&104.330&2.505&2.66$\pm 0.41$&2.40&2.80&2.30&2.56\\
$3^1S_0\rightarrow 2^1P_1\gamma$&106.260&2.644&-&2.10&-&1.7&2.60\\
$3^3S_1\rightarrow 1^3P_0\gamma$&521.46&1.152&0.054$\pm 0.08$&0.0001&0.03&0.01&0.15\\
$3^3S_1\rightarrow 1^3P_1\gamma$&476.799&0.641&0.018$\pm 0.001$&0.008&0.09&0.05&0.16\\
$3^3S_1\rightarrow 1^3P_2\gamma$&462.035&0.520&0.201$\pm 0.32$&0.015&0.13&0.45&0.0827\\
$3^1S_0\rightarrow 1^1P_1\gamma$&466.045&0.551&-&0.007&-&1.30&0.0084\\
\hline
\end{tabular*}
\end{table*}
 The M1 transitions contribute little to the total decay widths of bottomonium states and are weaker than the E1 transitions of bottomonium states. M1 transitions have been used to observe the spin-singlet states\cite{SG16}.
Allowed M1 transitions correspond to triplet-singlet transitions between S-wave states of the same n quantum number, while hindered M1 transitions are either triplet-singlet or singlet-triplet transitions between S-wave states of different n quantum numbers.\\

In order to calculate decay rates of hindered transitions we need to include relativistic corrections.There are three main types of corrections: relativistic modification of the non relativistic wave functions, relativistic modification of the electromagnetic transition operator, and finite-size corrections. In addition to these there are additional corrections arising from the quark
anomalous magnetic moment. Corrections to the wave function that give contributions to the transition amplitude are of two categories. Firstly from higher order potential corrections, which are distinguished as a) Zero recoil effect and b) recoil effects of the final state meson and secondly from colour octet effects. The colour octet effects are not included in potential model formulation and are not considered so far in radiative transitions.

 The M1 transition rates of bottomonium states have been calculated using equation (\ref{decay}). The resulting M1 radiative transition rates of these states are presented in Table \ref{M1}. In this table we give calculated values for decay rates of M1 radiative transition in comparison with the other relativistic and non relativistic quark models. We see from these results that the relativistic effects play a very important role in determining the bottomonium M1 transition rates. The relativistic effects reduce the decay rates of allowed transitions and increase the rates of hindered transitions. The M1 transition rates calculated in our model agree well with the values predicted by other theoretical models.
   
 \begin{table*}[h!]
\centering
\caption{\label{M1}\bf M1 transition rates of bottomonium.}
\setlength{\tabcolsep}{2pt}
\begin{tabular}{ccccccccc}
\hline
Transition&$k_0(MeV)$&$\Gamma$ eV&$\Gamma$ eV&$\Gamma^{NR}$ eV&$\Gamma$ eV&$\Gamma$ eV\\
&&This work&Experiment\cite{PDG}&Ref.\cite{DR03}&Ref.\cite{DR03}&Ref.\cite{SG16}&Ref.\cite{JS15}\\
\hline
$\Upsilon(1S)\rightarrow \eta_b(1S)\gamma$&60.833&10.65&-&12.2&9.7&10.0&9.34\\
$\Upsilon(2S)\rightarrow \eta_b(2S)\gamma$&23.352&0.603&-&1.50&1.6&0.59&0.58\\
$\Upsilon(3S)\rightarrow \eta_b(3S)\gamma$&16.027&0.195&-&0.8&0.9&0.25&0.658\\
$\Upsilon(2S)\rightarrow \eta_b(1S)\gamma$&605.291&31.47&12.47$\pm$4.90&1.3&1.3&81&56.5\\
$\eta_b(2S)\rightarrow \Upsilon(1S)\gamma$&525.752&12.16&-&2.5&2.4&68&45.0\\
$\Upsilon(3S)\rightarrow \eta_b(1S)\gamma$&944.887&3.56&10.36$\pm$1.70&3.1&2.5&60&57.0\\
$\eta_b(3S)\rightarrow \Upsilon(1S)\gamma$&874.828&1.60&-&7.1&5.8&74.0&51.0\\
$\Upsilon(3S)\rightarrow \eta_b(2S)\gamma$&383.528&4.47&<12.59$\pm$1.14&0.1&0.2&190&11.0\\
$\eta_b(3S)\rightarrow \Upsilon(2S)\gamma$&345.507&2.19&-&0.2&0.4&9.1&9.20\\
\hline
\end{tabular}
\end{table*}


\section{Conclusions}
\label{sec:C}
The bottomonium mass spectrum and electromagnetic transitions are investigated by adopting the quark-antiquark potential consisting of the confined one-gluon-exchange and the Lorentz scalar plus vector harmonic oscillator confinement potentials. We perform a nonperturbative calculations with Hamiltonian including spin-independent and dependent potentials.  \\

The bottomonium spectrum predicted by our quark model is in good agreement with the experimental data. We have calculated a large number of electromagnetic  decay widths showing
that our results are in reasonable agreement with the other model caculations in most of the cases.  The hyperfine mass splitting between the singlet and triplet 1S as well as 2S states are consistent with the experimental data and also with lattice QCD results. We have also predicted radially excited states of $\eta_b$ and their triplet states. The masses of P wave states calculated in our model are in good agreement with the experimental values and other theoretical models. Some of the higher excited states are 50-100 MeV heavier in our model. This may be due to the fact that coupled channel effects are significant in these states.\\

 Radiative decays are the dominant decay modes of the $b\bar{b}$ excited states having widths of about a fraction of MeV. In order to understand the $b\bar{b}$ spectrum and distinguish exotic states from conventional ones, it is very essential that the masses and the radiative decay widths of $b\bar{b}$ states are accurately determined. The calculated M1 transition rates reasonably agree with the other theoretical model predictions as listed in table \ref{M1}. It is clearly seen in this calculation that the relativistic effects play an important role in determining the M1 radiative transition rates, since the hindered transition rates are zero due to the wave function orthogonality in the NRQM formalism. The inclusion of
relativistic effects enhances the non-relativistically-hindered
radiative transition rates, predicting them larger than the allowed
ones by an order of magnitude. It is a good example for the importance
of relativity, even for some properties of heavy mesons.\\

\label{conc}
\begin{center}
\textbf{Acknowledgements}
\end{center}
One of the authors (APM) is grateful to BRNS,                                                             DAE, India for granting the project and JRF (37(3)/14/21/2014BRNS). 

\bibliography{mybib}

\begin{thebibliography}{10}
\expandafter\ifx\csname url\endcsname\relax
  \def\url#1{\texttt{#1}}\fi
\expandafter\ifx\csname urlprefix\endcsname\relax\def\urlprefix{URL }\fi
\expandafter\ifx\csname href\endcsname\relax
  \def\href#1#2{#2} \def\path#1{#1}\fi

\bibitem{HS77}
S.~W. Herb, et~al.,
  \href{http://link.aps.org/doi/10.1103/PhysRevLett.39.252}{Observation of a
  dimuon resonance at 9.5 gev in 400-gev proton-nucleus collisions}, Phys. Rev.
  Lett. 39 (1977) 252--255.
\newblock \href {http://dx.doi.org/10.1103/PhysRevLett.39.252}
  {\path{doi:10.1103/PhysRevLett.39.252}}.
\newline\urlprefix\url{http://link.aps.org/doi/10.1103/PhysRevLett.39.252}

\bibitem{WR77}
W.~R. Innes, et~al.,
  \href{http://link.aps.org/doi/10.1103/PhysRevLett.39.1240}{Observation of
  structure in the $\ensuremath{\Upsilon}$ region}, Phys. Rev. Lett. 39 (1977)
  1240--1242.
\newblock \href {http://dx.doi.org/10.1103/PhysRevLett.39.1240}
  {\path{doi:10.1103/PhysRevLett.39.1240}}.
\newline\urlprefix\url{http://link.aps.org/doi/10.1103/PhysRevLett.39.1240}

\bibitem{AB08}
B.~Aubert, et~al.,
  \href{http://link.aps.org/doi/10.1103/PhysRevLett.101.071801}{Observation of
  the bottomonium ground state in the decay
  $\ensuremath{\Upsilon}(3s)\ensuremath{\rightarrow}\ensuremath{\gamma}{\ensuremath{\eta}}_{b}$},
  Phys. Rev. Lett. 101 (2008) 071801.
\newblock \href {http://dx.doi.org/10.1103/PhysRevLett.101.071801}
  {\path{doi:10.1103/PhysRevLett.101.071801}}.
\newline\urlprefix\url{http://link.aps.org/doi/10.1103/PhysRevLett.101.071801}

\bibitem{BG10}
G.~Bonvicini, et~al.,
  \href{http://link.aps.org/doi/10.1103/PhysRevD.81.031104}{Measurement of the
  ${\ensuremath{\eta}}_{b}(1s)$ mass and the branching fraction for
  $\ensuremath{\Upsilon}(3s)\ensuremath{\rightarrow}\ensuremath{\gamma}{\ensuremath{\eta}}_{b}(1s)$},
  Phys. Rev. D 81 (2010) 031104.
\newblock \href {http://dx.doi.org/10.1103/PhysRevD.81.031104}
  {\path{doi:10.1103/PhysRevD.81.031104}}.
\newline\urlprefix\url{http://link.aps.org/doi/10.1103/PhysRevD.81.031104}

\bibitem{MR12}
R.~Mizuk, et~al.,
  \href{http://link.aps.org/doi/10.1103/PhysRevLett.109.232002}{Evidence for
  the ${\ensuremath{\eta}}_{b}(2s)$ and observation of
  ${h}_{b}(1p)\ensuremath{\rightarrow}{\ensuremath{\eta}}_{b}(1s)\ensuremath{\gamma}$
  and
  ${h}_{b}(2p)\ensuremath{\rightarrow}{\ensuremath{\eta}}_{b}(1s)\ensuremath{\gamma}$},
  Phys. Rev. Lett. 109 (2012) 232002.
\newblock \href {http://dx.doi.org/10.1103/PhysRevLett.109.232002}
  {\path{doi:10.1103/PhysRevLett.109.232002}}.
\newline\urlprefix\url{http://link.aps.org/doi/10.1103/PhysRevLett.109.232002}

\bibitem{SD12}
S.~Dobbs, et~al.,
  \href{http://link.aps.org/doi/10.1103/PhysRevLett.109.082001}{Observation of
  the ${\ensuremath{\eta}}_{b}(2s)$ meson in
  $\ensuremath{\Upsilon}(2s)\ensuremath{\rightarrow}\ensuremath{\gamma}{\ensuremath{\eta}}_{b}(2s)$,
  ${\ensuremath{\eta}}_{b}(2s)\ensuremath{\rightarrow}$ hadrons and
  confirmation of the ${\ensuremath{\eta}}_{b}(1s)$ meson}, Phys. Rev. Lett.
  109 (2012) 082001.
\newblock \href {http://dx.doi.org/10.1103/PhysRevLett.109.082001}
  {\path{doi:10.1103/PhysRevLett.109.082001}}.
\newline\urlprefix\url{http://link.aps.org/doi/10.1103/PhysRevLett.109.082001}

\bibitem{LJ12}
J.~P. Lees, et~al.,
  \href{http://link.aps.org/doi/10.1103/PhysRevD.84.072002}{Study of radiative
  bottomonium transitions using converted photons}, Phys. Rev. D 84 (2011)
  072002.
\newblock \href {http://dx.doi.org/10.1103/PhysRevD.84.072002}
  {\path{doi:10.1103/PhysRevD.84.072002}}.
\newline\urlprefix\url{http://link.aps.org/doi/10.1103/PhysRevD.84.072002}

\bibitem{AI12}
I.~Adachi, et~al.,
  \href{http://link.aps.org/doi/10.1103/PhysRevLett.108.032001}{First
  observation of the $p$-wave spin-singlet bottomonium states ${h}_{b}(1p)$ and
  ${h}_{b}(2p)$}, Phys. Rev. Lett. 108 (2012) 032001.
\newblock \href {http://dx.doi.org/10.1103/PhysRevLett.108.032001}
  {\path{doi:10.1103/PhysRevLett.108.032001}}.
\newline\urlprefix\url{http://link.aps.org/doi/10.1103/PhysRevLett.108.032001}

\bibitem{KH83}
K.~Han, et~al.,
  \href{http://link.aps.org/doi/10.1103/PhysRevLett.49.1612}{Observation of
  $p$-wave $b\overline{b}$ bound states}, Phys. Rev. Lett. 49 (1982)
  1612--1616.
\newblock \href {http://dx.doi.org/10.1103/PhysRevLett.49.1612}
  {\path{doi:10.1103/PhysRevLett.49.1612}}.
\newline\urlprefix\url{http://link.aps.org/doi/10.1103/PhysRevLett.49.1612}

\bibitem{EG82}
G.~Eigen, et~al.,
  \href{http://link.aps.org/doi/10.1103/PhysRevLett.49.1616}{Evidence for
  $\chi_b'$ production in the exclusive reaction $\upsilon" \rightarrow
  \gamma\chi_b'\rightarrow\gamma\gamma\upsilon' ~\rm{or}
  \ensuremath{\gamma}\ensuremath{\gamma}\ensuremath{\Upsilon})$}, Phys. Rev.
  Lett. 49 (1982) 1616--1619.
\newblock \href {http://dx.doi.org/10.1103/PhysRevLett.49.1616}
  {\path{doi:10.1103/PhysRevLett.49.1616}}.
\newline\urlprefix\url{http://link.aps.org/doi/10.1103/PhysRevLett.49.1616}

\bibitem{CK83}
C.~Klopfenstein, et~al.,
  \href{http://link.aps.org/doi/10.1103/PhysRevLett.51.160}{Observation of the
  lowest $p$-wave $b\overline{b}$ bound states}, Phys. Rev. Lett. 51 (1983)
  160--163.
\newblock \href {http://dx.doi.org/10.1103/PhysRevLett.51.160}
  {\path{doi:10.1103/PhysRevLett.51.160}}.
\newline\urlprefix\url{http://link.aps.org/doi/10.1103/PhysRevLett.51.160}

\bibitem{FH83}
F.~Pauss, et~al.,
  \href{http://www.sciencedirect.com/science/article/pii/0370269383915393}{Observation
  of χb production in the exclusive reaction $\upsilon '\to \gamma\gamma\chi_b
  \to\gamma\gamma\upsilon\to\gamma\gamma (e^+e^-~\rm{or}\mu^+\mu^-)$}, Physics
  Letters B 130~(6) (1983) 439 -- 443.
\newblock \href
  {http://dx.doi.org/http://dx.doi.org/10.1016/0370-2693(83)91539-3}
  {\path{doi:http://dx.doi.org/10.1016/0370-2693(83)91539-3}}.
\newline\urlprefix\url{http://www.sciencedirect.com/science/article/pii/0370269383915393}

\bibitem{GA12}
G.~Aad, et~al.,
  \href{http://link.aps.org/doi/10.1103/PhysRevLett.108.152001}{Observation of
  a new ${\ensuremath{\chi}}_{b}$ state in radiative transitions to
  $\ensuremath{\Upsilon}(1s)$ and $\ensuremath{\Upsilon}(2s)$ at atlas}, Phys.
  Rev. Lett. 108 (2012) 152001.
\newblock \href {http://dx.doi.org/10.1103/PhysRevLett.108.152001}
  {\path{doi:10.1103/PhysRevLett.108.152001}}.
\newline\urlprefix\url{http://link.aps.org/doi/10.1103/PhysRevLett.108.152001}

\bibitem{VM12}
V.~M. Abazov, et~al.,
  \href{http://link.aps.org/doi/10.1103/PhysRevD.86.031103}{Observation of a
  narrow mass state decaying into
  $\ensuremath{\Upsilon}(1s)+\ensuremath{\gamma}$ in $p\overline{p}$ collisions
  at $\sqrt{s}=1.96\text{ }\text{ }\mathrm{TeV}$}, Phys. Rev. D 86 (2012)
  031103.
\newblock \href {http://dx.doi.org/10.1103/PhysRevD.86.031103}
  {\path{doi:10.1103/PhysRevD.86.031103}}.
\newline\urlprefix\url{http://link.aps.org/doi/10.1103/PhysRevD.86.031103}

\bibitem{AR14}
R.~Aaij, et~al., \href{http://dx.doi.org/10.1007/JHEP10(2014)088}{Measurement
  of the $\chi_b$(3p) mass and of the relative rate of $\chi_{b1}(1p)$ and
  $\chi_{b2}(1p)$ production}, Journal of High Energy Physics 2014~(10) (2014)
  88.
\newblock \href {http://dx.doi.org/10.1007/JHEP10(2014)088}
  {\path{doi:10.1007/JHEP10(2014)088}}.
\newline\urlprefix\url{http://dx.doi.org/10.1007/JHEP10(2014)088}

\bibitem{GB04}
G.~Bonvicini, et~al.,
  \href{http://link.aps.org/doi/10.1103/PhysRevD.70.032001}{First observation
  of a $\ensuremath{\Upsilon}(1d)$ state}, Phys. Rev. D 70 (2004) 032001.
\newblock \href {http://dx.doi.org/10.1103/PhysRevD.70.032001}
  {\path{doi:10.1103/PhysRevD.70.032001}}.
\newline\urlprefix\url{http://link.aps.org/doi/10.1103/PhysRevD.70.032001}

\bibitem{PD10}
P.~del Amo~Sanchez, et~al.,
  \href{http://link.aps.org/doi/10.1103/PhysRevD.82.111102}{Observation of the
  $\mathit{\ensuremath{\Upsilon}}({1}^{3}{D}_{J})$ bottomonium state through
  decays to
  ${\ensuremath{\pi}}^{+}{\ensuremath{\pi}}^{\ensuremath{-}}\mathit{\ensuremath{\Upsilon}}(1s)$},
  Phys. Rev. D 82 (2010) 111102.
\newblock \href {http://dx.doi.org/10.1103/PhysRevD.82.111102}
  {\path{doi:10.1103/PhysRevD.82.111102}}.
\newline\urlprefix\url{http://link.aps.org/doi/10.1103/PhysRevD.82.111102}

\bibitem{BD85}
D.~Besson, et~al.,
  \href{http://link.aps.org/doi/10.1103/PhysRevLett.54.381}{Observation of new
  structure in the ${e}^{+}{e}^{\ensuremath{-}}$ cross section above the
  $\ensuremath{\Upsilon}(4s)$}, Phys. Rev. Lett. 54 (1985) 381--384.
\newblock \href {http://dx.doi.org/10.1103/PhysRevLett.54.381}
  {\path{doi:10.1103/PhysRevLett.54.381}}.
\newline\urlprefix\url{http://link.aps.org/doi/10.1103/PhysRevLett.54.381}

\bibitem{LD85}
D.~M.~J. Lovelock, et~al.,
  \href{http://link.aps.org/doi/10.1103/PhysRevLett.54.377}{Masses, widths, and
  leptonic widths of the higher upsilon resonances}, Phys. Rev. Lett. 54 (1985)
  377--380.
\newblock \href {http://dx.doi.org/10.1103/PhysRevLett.54.377}
  {\path{doi:10.1103/PhysRevLett.54.377}}.
\newline\urlprefix\url{http://link.aps.org/doi/10.1103/PhysRevLett.54.377}

\bibitem{AB12}
A.~Bondar, et~al.,
  \href{http://link.aps.org/doi/10.1103/PhysRevLett.108.122001}{Observation of
  two charged bottomoniumlike resonances in $\ensuremath{\Upsilon}(5s)$
  decays}, Phys. Rev. Lett. 108 (2012) 122001.
\newblock \href {http://dx.doi.org/10.1103/PhysRevLett.108.122001}
  {\path{doi:10.1103/PhysRevLett.108.122001}}.
\newline\urlprefix\url{http://link.aps.org/doi/10.1103/PhysRevLett.108.122001}

\bibitem{HG90}
H.~Georgi,
  \href{http://www.sciencedirect.com/science/article/pii/037026939091128X}{An
  effective field theory for heavy quarks at low energies}, Physics Letters B
  240~(3) (1990) 447 -- 450.
\newblock \href
  {http://dx.doi.org/http://dx.doi.org/10.1016/0370-2693(90)91128-X}
  {\path{doi:http://dx.doi.org/10.1016/0370-2693(90)91128-X}}.
\newline\urlprefix\url{http://www.sciencedirect.com/science/article/pii/037026939091128X}

\bibitem{M97}
A.~V. Manohar, \href{http://dx.doi.org/10.1007/BFb0104294}{Effective field
  theories}, Springer Berlin Heidelberg, Berlin, Heidelberg, 1997, pp.
  311--362.
\newblock \href {http://dx.doi.org/10.1007/BFb0104294}
  {\path{doi:10.1007/BFb0104294}}.
\newline\urlprefix\url{http://dx.doi.org/10.1007/BFb0104294}

\bibitem{NB05}
N.~Brambilla, A.~Pineda, J.~Soto, A.~Vairo,
  \href{https://link.aps.org/doi/10.1103/RevModPhys.77.1423}{Effective-field
  theories for heavy quarkonium}, Rev. Mod. Phys. 77 (2005) 1423--1496.
\newblock \href {http://dx.doi.org/10.1103/RevModPhys.77.1423}
  {\path{doi:10.1103/RevModPhys.77.1423}}.
\newline\urlprefix\url{https://link.aps.org/doi/10.1103/RevModPhys.77.1423}

\bibitem{PC92}
P.~C. Vinodkumar, K.~B. Vijaya~Kumar, S.~B. Khadkikar, Effect of the confined
  gluons in quark-quark interaction, Pramana 39~(1) (1992) 47--70.
\newblock \href {http://dx.doi.org/10.1007/BF02853034}
  {\path{doi:10.1007/BF02853034}}.

\bibitem{SB91}
S.~B. Khadkikar, K.~B.~V. Kumar, \{NN\} scattering with exchange of confined
  gluons, Phys. Lett. B 254~(3) (1991) 320 -- 324.
\newblock \href {http://dx.doi.org/10.1016/0370-2693(91)91162-O}
  {\path{doi:10.1016/0370-2693(91)91162-O}}.

\bibitem{KB93}
K.~B. Vijaya~Kumar, S.~B. Khadkikar, \{NN\} interaction in a relativistic
  harmonic model with confined gluons, Nucl. Phys. A 556~(3) (1993) 396 -- 408.
\newblock \href {http://dx.doi.org/10.1016/0375-9474(93)90368-8}
  {\path{doi:10.1016/0375-9474(93)90368-8}}.

\bibitem{KB97}
K.~B. Vijaya~Kumar, A.~K. Rath, S.~B. Khadkikar, Meson spectroscopy with
  confined one gluon exchange potential, Pramana 48~(5) (1997) 997--1004.
\newblock \href {http://dx.doi.org/10.1007/BF02847459}
  {\path{doi:10.1007/BF02847459}}.

\bibitem{SB83}
S.~B. Khadkikar, S.~K. Gupta, Magnetic moments of light baryons in harmonic
  model, Phys. Lett. B 124~(6) (1983) 523 -- 526.
\newblock \href {http://dx.doi.org/.10.1016/0370-2693(83)91566-6}
  {\path{doi:.10.1016/0370-2693(83)91566-6}}.

\bibitem{VH04}
K.~B. Vijaya~Kumar, B.~Hanumaiah, S.~Pepin, Meson spectrum in a relativistic
  harmonic model with instanton-induced interaction, Eur.Phys. J A 19~(2)
  (2004) 247--250.
\newblock \href {http://dx.doi.org/10.1140/epja/i2002-10287-1}
  {\path{doi:10.1140/epja/i2002-10287-1}}.

\bibitem{VB09}
K.~B. Vijaya~Kumar, Bhavyashri, Y.-L. Ma, A.~P. Monteiro, P wave meson spectrum
  in a relativistic model with instanton induced interaction, Int. J. Mod.
  Phys. A 22 (2009) 4209.
\newblock \href {http://arxiv.org/abs/arXiv:0811.4308 [hep-ph]}
  {\path{arXiv:arXiv:0811.4308 [hep-ph]}}.

\bibitem{GG57}
S.~Gartenhaus, C.~Schwartz, Center-of-mass motion in many-particle systems,
  Phys. Rev. 108 (1957) 482--490.
\newblock \href {http://dx.doi.org/10.1103/PhysRev.108.482}
  {\path{doi:10.1103/PhysRev.108.482}}.

\bibitem{WJ88}
W.~Kwong, J.~L. Rosner, \textit{D}-wave quarkonium levels of the
  \textit{\ensuremath{\Upsilon}} family, Phys. Rev. D 38 (1988) 279--297.
\newblock \href {http://dx.doi.org/10.1103/PhysRevD.38.279}
  {\path{doi:10.1103/PhysRevD.38.279}}.

\bibitem{NL78}
V.~Novikov, L.~Okun, M.~Shifman, A.~Vainshtein, M.~Voloshin, V.~Zakharov,
  Charmonium and gluons, Phys. Rep. 41~(1) (1978) 1 -- 133.
\newblock \href {http://dx.doi.org/10.1016/0370-1573(78)90120-5}
  {\path{doi:10.1016/0370-1573(78)90120-5}}.

\bibitem{CE05}
N.~Brambilla, et~al., Heavy quarkonium physics, CERN Yellow Report,
  CERN-2005-005\href {http://arxiv.org/abs/hep-ph/0412158}
  {\path{arXiv:hep-ph/0412158}}.

\bibitem{WE86}
W.~Caswell, G.~Lepage, Effective lagrangians for bound state problems in qed,
  qcd, and other field theories, Phys. Lett. B 167~(4) (1986) 437 -- 442.
\newblock \href {http://dx.doi.org/10.1016/0370-2693(86)91297-9}
  {\path{doi:10.1016/0370-2693(86)91297-9}}.

\bibitem{BB95}
G.~T. Bodwin, E.~Braaten, G.~P. Lepage, Rigorous qcd analysis of inclusive
  annihilation and production of heavy quarkonium, Phys. Rev. D 51 (1995)
  1125--1171.
\newblock \href {http://arxiv.org/abs/hep-ph/9407339}
  {\path{arXiv:hep-ph/9407339}}.

\bibitem{NB99}
N.~Brambilla, A.~Pineda, J.~Soto, A.~Vairo, The heavy quarkonium spectrum at
  order m$\alpha^5_s\ln\alpha_s$, Phys. Lett. B 470~(1–4) (1999) 215 -- 222.
\newblock \href {http://arxiv.org/abs/hep-ph/9910238}
  {\path{arXiv:hep-ph/9910238}}.

\bibitem{BA96}
B.~A. Thacker, G.~P. Lepage, Heavy-quark bound states in lattice qcd, Phys.
  Rev. D 43 (1991) 196--208.
\newblock \href {http://dx.doi.org/10.1103/PhysRevD.43.196}
  {\path{doi:10.1103/PhysRevD.43.196}}.

\bibitem{FD69}
D.~Faiman, A.~W. Hendry,
  \href{http://link.aps.org/doi/10.1103/PhysRev.180.1572}{Electromagnetic
  decays of baryon resonances in the harmonic-oscillator model}, Phys. Rev. 180
  (1969) 1572--1577.
\newblock \href {http://dx.doi.org/10.1103/PhysRev.180.1572}
  {\path{doi:10.1103/PhysRev.180.1572}}.
\newline\urlprefix\url{http://link.aps.org/doi/10.1103/PhysRev.180.1572}

\bibitem{RP71}
R.~P. Feynman, M.~Kislinger, F.~Ravndal,
  \href{http://link.aps.org/doi/10.1103/PhysRevD.3.2706}{Current matrix
  elements from a relativistic quark model}, Phys. Rev. D 3 (1971) 2706--2732.
\newblock \href {http://dx.doi.org/10.1103/PhysRevD.3.2706}
  {\path{doi:10.1103/PhysRevD.3.2706}}.
\newline\urlprefix\url{http://link.aps.org/doi/10.1103/PhysRevD.3.2706}

\bibitem{KS89}
K.~Shimizu, \href{http://stacks.iop.org/0034-4885/52/i=1/a=001}{Study of
  baryon-baryon interactions and nuclear properties in the quark cluster
  model}, Reports on Progress in Physics 52~(1) (1989) 1.
\newline\urlprefix\url{http://stacks.iop.org/0034-4885/52/i=1/a=001}

\bibitem{FA83}
A.~Faessler, F.~Fernandez, G.~Lübeck, K.~Shimizu,
  \href{http://www.sciencedirect.com/science/article/pii/0375947483902191}{The
  nucleon-nucleon interaction and the role of the [42] orbital six-quark
  symmetry}, Nuclear Physics A 402~(3) (1983) 555 -- 568.
\newblock \href
  {http://dx.doi.org/http://dx.doi.org/10.1016/0375-9474(83)90219-1}
  {\path{doi:http://dx.doi.org/10.1016/0375-9474(83)90219-1}}.
\newline\urlprefix\url{http://www.sciencedirect.com/science/article/pii/0375947483902191}

\bibitem{SN86}
S.~Capstick, N.~Isgur, Baryons in a relativized quark model with
  chromodynamics, Phys. Rev. D 34 (1986) 2809--2835.
\newblock \href {http://dx.doi.org/10.1103/PhysRevD.34.2809}
  {\path{doi:10.1103/PhysRevD.34.2809}}.

\bibitem{IM92}
I.~M. Narodetskii, R.~Ceuleneer, C.~Semay, Hyperfine interaction in the
  nonrelativistic quark model and the convergence of harmonic oscillator
  variational method, Journal of Physics G: Nuclear and Particle Physics
  18~(12) (1992) 1901.

\bibitem{SB85}
S.~B. Khadkikar, \href{http://dx.doi.org/10.1007/BF02894818}{Relativistic
  harmonic confinement of quarks and gluons}, Pramana 24~(1) (1985) 63--68.
\newblock \href {http://dx.doi.org/10.1007/BF02894818}
  {\path{doi:10.1007/BF02894818}}.
\newline\urlprefix\url{http://dx.doi.org/10.1007/BF02894818}

\bibitem{SB87}
S.~B. Khadkikar, P.~C. Vinodkumar,
  \href{http://dx.doi.org/10.1007/BF02845677}{Confinement models for gluons},
  Pramana 29~(1) (1987) 39--52.
\newblock \href {http://dx.doi.org/10.1007/BF02845677}
  {\path{doi:10.1007/BF02845677}}.
\newline\urlprefix\url{http://dx.doi.org/10.1007/BF02845677}

\bibitem{AM80}
A.~Martin, A fit of upsilon and charmonium spectra, Phys. Lett. B 93~(3) (1980)
  338 -- 342.
\newblock \href {http://dx.doi.org/10.1016/0370-2693(80)90527-4}
  {\path{doi:10.1016/0370-2693(80)90527-4}}.

\bibitem{EJ78}
E.~Eichten, K.~Gottfried, T.~Kinoshita, K.~D. Lane, T.~M. Yan, Charmonium: The
  model, Phys. Rev. D 17 (1978) 3090--3117.
\newblock \href {http://dx.doi.org/10.1103/PhysRevD.17.3090}
  {\path{doi:10.1103/PhysRevD.17.3090}}.

\bibitem{WS81}
W.~Buchm\"uller, S.~H.~H. Tye, Quarkonia and quantum chromodynamics, Phys. Rev.
  D 24 (1981) 132--156.
\newblock \href {http://dx.doi.org/10.1103/PhysRevD.24.132}
  {\path{doi:10.1103/PhysRevD.24.132}}.

\bibitem{CJ77}
C.~Quigg, J.~L. Rosner, Quarkonium level spacings, Phys. Lett. B 71~(1) (1977)
  153 -- 157.
\newblock \href {http://dx.doi.org/10.1016/0370-2693(77)90765-1}
  {\path{doi:10.1016/0370-2693(77)90765-1}}.

\bibitem{DR03}
D.~Ebert, R.~N. Faustov, V.~O. Galkin, Properties of heavy quarkonia and
  ${B}_{c}$ mesons in the relativistic quark model, Phys. Rev. D 67 (2003)
  014027.
\newblock \href {http://arxiv.org/abs/hep-ph/0210381v2}
  {\path{arXiv:hep-ph/0210381v2}}.

\bibitem{SN85}
S.~Godfrey, N.~Isgur, Mesons in a relativized quark model with chromodynamics,
  Phys. Rev. D 32 (1985) 189--231.
\newblock \href {http://dx.doi.org/10.1103/PhysRevD.32.189}
  {\path{doi:10.1103/PhysRevD.32.189}}.

\bibitem{AA05}
A.~Abd El-Hady, J.~R. Spence, J.~P. Vary, Radiative decays of ${B}_{c}$ mesons
  in a bethe-salpeter model, Phys. Rev. D 71 (2005) 034006.
\newblock \href {http://arxiv.org/abs/hep-ph/0603139v1}
  {\path{arXiv:hep-ph/0603139v1}}.

\bibitem{EC94}
E.~J. Eichten, C.~Quigg, Mesons with beauty and charm: Spectroscopy, Phys. Rev.
  D 49 (1994) 5845--5856.
\newblock \href {http://arxiv.org/abs/hep-ph/9402210}
  {\path{arXiv:hep-ph/9402210}}.

\bibitem{SA95}
S.~S. Gershtein, V.~V. Kiselev, A.~K. Likhoded, A.~V. Tkabladze, Physics of bc
  -mesons, Physics-Uspekhi 38~(1) (1995) 1.
\newblock \href {http://arxiv.org/abs/hep-ph/9504319}
  {\path{arXiv:hep-ph/9504319}}.

\bibitem{BK05}
Bhavyashri, K.~B.~V. Kumar, B.~Hanumaiah, S.~Sarangi, S.-G. Zhou,
  \href{http://stacks.iop.org/0954-3899/31/i=8/a=026}{Meson spectrum in a
  non-relativistic model with instanton-induced interaction}, Journal of
  Physics G: Nuclear and Particle Physics 31~(8) (2005) 981.
\newline\urlprefix\url{http://stacks.iop.org/0954-3899/31/i=8/a=026}

\bibitem{PDG}
K.~Olive, et~al., \href{http://stacks.iop.org/1674-1137/38/i=9/a=090001}{Review
  of particle physics}, Chinese Physics C 38~(9) (2014) 090001.
\newline\urlprefix\url{http://stacks.iop.org/1674-1137/38/i=9/a=090001}

\bibitem{SM10}
S.~Meinel,
  \href{http://link.aps.org/doi/10.1103/PhysRevD.82.114502}{Bottomonium
  spectrum at order ${v}^{6}$ from domain-wall lattice qcd: Precise results for
  hyperfine splittings}, Phys. Rev. D 82 (2010) 114502.
\newblock \href {http://dx.doi.org/10.1103/PhysRevD.82.114502}
  {\path{doi:10.1103/PhysRevD.82.114502}}.
\newline\urlprefix\url{http://link.aps.org/doi/10.1103/PhysRevD.82.114502}

\bibitem{BA04}
B.~A. Kniehl, A.~A. Penin, A.~Pineda, V.~A. Smirnov, M.~Steinhauser,
  \href{http://link.aps.org/doi/10.1103/PhysRevLett.92.242001}{Mass of the
  ${\ensuremath{\eta}}_{b}$ and ${\ensuremath{\alpha}}_{s}$ from the
  nonrelativistic renormalization group}, Phys. Rev. Lett. 92 (2004) 242001.
\newblock \href {http://dx.doi.org/10.1103/PhysRevLett.92.242001}
  {\path{doi:10.1103/PhysRevLett.92.242001}}.
\newline\urlprefix\url{http://link.aps.org/doi/10.1103/PhysRevLett.92.242001}

\bibitem{JS15}
J.~Segovia, P.~G. Ortega, D.~R. Entem, F.~Fern\'andez,
  \href{http://link.aps.org/doi/10.1103/PhysRevD.93.074027}{Bottomonium
  spectrum revisited}, Phys. Rev. D 93 (2016) 074027.
\newblock \href {http://dx.doi.org/10.1103/PhysRevD.93.074027}
  {\path{doi:10.1103/PhysRevD.93.074027}}.
\newline\urlprefix\url{http://link.aps.org/doi/10.1103/PhysRevD.93.074027}

\bibitem{SG16}
S.~Godfrey, K.~Moats,
  \href{http://link.aps.org/doi/10.1103/PhysRevD.92.054034}{Bottomonium mesons
  and strategies for their observation}, Phys. Rev. D 92 (2015) 054034.
\newblock \href {http://dx.doi.org/10.1103/PhysRevD.92.054034}
  {\path{doi:10.1103/PhysRevD.92.054034}}.
\newline\urlprefix\url{http://link.aps.org/doi/10.1103/PhysRevD.92.054034}

\bibitem{JF13}
J.~Ferretti, E.~Santopinto,
  \href{http://link.aps.org/doi/10.1103/PhysRevD.90.094022}{Higher mass
  bottomonia}, Phys. Rev. D 90 (2014) 094022.
\newblock \href {http://dx.doi.org/10.1103/PhysRevD.90.094022}
  {\path{doi:10.1103/PhysRevD.90.094022}}.
\newline\urlprefix\url{http://link.aps.org/doi/10.1103/PhysRevD.90.094022}

\bibitem{Li12}
J.-F. Liu, G.-J. Ding,
  \href{http://dx.doi.org/10.1140/epjc/s10052-012-1981-6}{Bottomonium spectrum
  with coupled-channel effects}, The European Physical Journal C 72~(4) (2012)
  1--15.
\newblock \href {http://dx.doi.org/10.1140/epjc/s10052-012-1981-6}
  {\path{doi:10.1140/epjc/s10052-012-1981-6}}.
\newline\urlprefix\url{http://dx.doi.org/10.1140/epjc/s10052-012-1981-6}

\bibitem{YL16}
Y.~Lu, M.~N. Anwar, B.-S. Zou,
  \href{http://link.aps.org/doi/10.1103/PhysRevD.94.034021}{Coupled-channel
  effects for the bottomonium with realistic wave functions}, Phys. Rev. D 94
  (2016) 034021.
\newblock \href {http://dx.doi.org/10.1103/PhysRevD.94.034021}
  {\path{doi:10.1103/PhysRevD.94.034021}}.
\newline\urlprefix\url{http://link.aps.org/doi/10.1103/PhysRevD.94.034021}

\bibitem{EG08}
E.~Eichten, S.~Godfrey, H.~Mahlke, J.~L. Rosner,
  \href{https://link.aps.org/doi/10.1103/RevModPhys.80.1161}{Quarkonia and
  their transitions}, Rev. Mod. Phys. 80 (2008) 1161--1193.
\newblock \href {http://dx.doi.org/10.1103/RevModPhys.80.1161}
  {\path{doi:10.1103/RevModPhys.80.1161}}.
\newline\urlprefix\url{https://link.aps.org/doi/10.1103/RevModPhys.80.1161}

\bibitem{GR92}
A.~K. Grant, J.~L. Rosner,
  \href{https://link.aps.org/doi/10.1103/PhysRevD.46.3862}{Dipole transition
  matrix elements for systems with power-law potentials}, Phys. Rev. D 46
  (1992) 3862--3870.
\newblock \href {http://dx.doi.org/10.1103/PhysRevD.46.3862}
  {\path{doi:10.1103/PhysRevD.46.3862}}.
\newline\urlprefix\url{https://link.aps.org/doi/10.1103/PhysRevD.46.3862}

\bibitem{SR07}
S.~F. Radford, W.~W. Repko,
  \href{http://link.aps.org/doi/10.1103/PhysRevD.75.074031}{Potential model
  calculations and predictions for heavy quarkonium}, Phys. Rev. D 75 (2007)
  074031.
\newblock \href {http://dx.doi.org/10.1103/PhysRevD.75.074031}
  {\path{doi:10.1103/PhysRevD.75.074031}}.
\newline\urlprefix\url{http://link.aps.org/doi/10.1103/PhysRevD.75.074031}

\end{thebibliography}

\end{document}